\documentclass[12pt]{article}

\begin{document}

\title{\bf Nonlinear gauge interactions: a possible
solution to the ``measurement problem" in quantum mechanics}

\date{}

\author{Johan Hansson
\footnote{Department of Physics, Lule{\aa} University of
Technology, SE-971 87 Lule\aa, Sweden}}

\maketitle

\begin{abstract}
Two fundamental, and unsolved problems in physics are:
\\
i) the resolution of the ``measurement problem" in quantum
mechanics
\\
ii) the quantization of strongly nonlinear (nonabelian) gauge
theories. The aim of this paper is to suggest that these two
problems might be linked, and that a mutual, simultaneous solution
to both might exist.

We propose that the mechanism responsible for the ``collapse of
the wave function" in quantum mechanics is the nonlinearities
already present in the theory via nonabelian gauge interactions.
Unlike all other models of spontaneous collapse, our proposal is,
to the best of our knowledge, the only one which does not
introduce any new elements into the theory.
A possible experimental test of the model would be to compare the
coherence lengths - here defined as the distance over which
quantum mechanical superposition is still valid - for,
\textit{e.g.}, electrons and photons in a double-slit experiment.
The electrons should have a finite coherence length, while photons
should have a much longer coherence length (in principle infinite,
if gravity - a very weak effect indeed unless we approach the
Planck scale - is ignored).
\\
\\
\end{abstract}

\newpage

\section{Introduction}
Simply stated, the ``measurement problem" in quantum mechanics is
that measuring instruments, and observers, are made up of quantum
entities (atoms, etc) and, according to the Schr\"{o}dinger
equation, quantum mechanical superpositions never stop. How then,
is a definite outcome of a measured ``observable", instead of the
predicted quantum superposition, possible? That is, where, and
how, is the superposition broken by a non-unitary ``collapse"? The
orthodox theory is completely silent on this point, and it is
simply postulated that somewhere between observed (quantum) object
and (classical) observer, the ``collapse" takes place.

As a possible solution to this, we start by noting that:
\begin{itemize}
  \item The measurement problem, or the collapse of the
  wavefunction in quantum mechanics is not solved.
  \item The quantization of \textit{strongly} nonlinear gauge
 theories is not solved, \textit{e.g.} strongly coupled
 Yang-Mills and gravity near the Planck-scale.
\end{itemize}
These statements can essentially be rephrased as:
\begin{itemize}
  \item The absolute backbone of quantum mechanics is the
  \textit{superposition principle} \cite{Dirac}
  (\textit{e.g.}, interfering
  amplitudes, summation of Feynman diagrams, etc). It is also
  well known that superposition requires \textit{linear} equations
  , \textit{i.e.}, the sum of two different solutions to a nonlinear
  equation is generally \textit{not} a solution, ruining superposition.
  The Hilbert space of quantum mechanics and the Fock space of
  quantum field theory are \textit{linear} spaces (based on the
  superposition requirement), suitable for linear mappings or
  operators.
  \item Nonabelian, \textit{i.e.} noncommutative, gauge field theories describing the fundamental
  interactions obey \textit{nonlinear}
  evolution equations in the gauge fields. As the gauge fields are
  supposed to
  be quantum mechanical this is in apparent
  contradiction to the previous statement. For convenience, we will
  write down the
  evolution equations for (pure) Yang-Mills fields below. Although
  the fermion evolution obeys linear equations, it
  becomes ``contaminated"
  by nonlinearities through the interactions.
\end{itemize}

The nonabelian vector gauge fields, $\mathbf{A}$, are governed by
a set of coupled, second order, nonlinear PDEs on Minkowski
spacetime. (The general argument for gravity is similar
\cite{Penrose}, but involve tensor fields constituting a dynamical
spacetime. We do not explicitly write down those equations.) For
pure Yang-Mills the evolution equations are given by
\begin{equation}
(\partial^{\mu} - g [A^{\mu}, \, ])_a ^b (\partial_{\mu} A_{\nu} -
\partial_{\nu} A_{\mu} -g[A_{\mu}, A_{\nu}])_b = 0 ,
\end{equation}
where $g$ is the coupling constant and $a$, $b$ are indices of the
gauge group, \textit{e.g.}, $a, b \in 1,2,3$ for $SU(2)$ and $a, b
\in 1,...,8$ for $SU(3)$. Summation over repeated indices is
implied. The operator (``covariant derivative") at the left works
according to $(\partial^{\mu} - g [A^{\mu}, \, ])(anything) =
\partial^{\mu}(anything)- g [A^{\mu}, anything]$. We see that we
get highly nonlinear (quadratic and cubic) terms in the gauge
fields, especially when the coupling constant, $g$, is large. The
commutator terms, ``square brackets", vanish identically for
abelian fields (\textit{e.g.} photons) as the gauge fields then
commute, leaving only the ordinary, linear Maxwell equations.

 In the Feynman path-integral formulation of quantum
mechanics \cite{Feynman} the nonlinearities can be ``hidden" in
the action functional, but as the Schr\"{o}dinger, Heisenberg and
path-integral formulations are equivalent, a problem in one of
them must translate into a problem in all.

That quantization of nonabelian theories is troublesome, and
hitherto unsolved in the strong field limit, or equivalently in
the large coupling constant limit, arises from their nonlinear
form. (One example of this, among many, is the ``Gribov ambiguity"
\cite{Gribov}, which implies non-uniqueness of gauge
transformations. This effect is absent in perturbation theory.)
The ``ghost" fields needed to preserve unitarity in nonabelian
quantum field theory might be seen as a consequence of trying to
constrain the truly nonlinear theory into a mildly nonlinear
theory defined by Feynman diagrams.

Also, the intuitively simple picture of elementary particles
(quanta!), which arises from perturbation theory, \textit{e.g.}
through Feynman diagrams, is absent in the strongly coupled
regime, and if we assume that ``quantization" is equivalent to the
exchange of elementary quanta, it is obvious that quantum
mechanics somehow must fail in the strong regime.

\section{Idea}
Instead of trying to solve these two problems separately, we
propose that they are, in fact, related. We assume that instead of
quantizing strongly nonlinear theories, it is their actual
nonlinearities which are responsible for breaking quantum
mechanical superposition, hence ``turning a vice into a virtue".

We thus get a \textit{self-induced collapse} into the ordinary
world of chairs, tables, people and indeed also recorded
elementary particle tracks in a photographic emulsion, a bubble
chamber, or a modern multi-purpose computer-aided detector.

It should be stated that this is quite the opposite of what is
normally considered in so called ``quantum chaos". There one
studies what consequences a chaotic classical system has for the
corresponding (chaos suppressing) quantum system, especially in
the semi-classical, short wavelength regime. We, on the other
hand, propose that the nonlinearity in the underlying dynamics
could be responsible for the seemingly random character of quantum
mechanics. This is possible as it is well known that classical
gauge field theories can be chaotic \cite{Biro} (in which case no
analytic, closed formula solutions exist, seemingly precluding any
simple ``quantization recipe"). In ``usual", non-relativistic
quantum chaos the nonlinearities would merely introduce higher
order terms in the operator potentials in the Hamiltonian.
However, it should be remembered that non-relativistic quantum
mechanics is but an approximation to relativistic quantum
mechanics, which itself is an approximation to quantum field
theory. In quantum field theory the conceptual difference between
fermions (``matter", $\psi$) and bosons (``interactions") almost
disappears. \textit{Both} fermions and bosons are now operators,
differing only in that the former obey anti-commutation relations,
while the latter obey commutation relations. Hence $\psi$ is no
longer privileged as is the case in non-relativistic quantum
mechanics.

Also, from an intuitive physical viewpoint, it is obvious that
nonabelian gauge bosons cannot behave like their abelian
counterparts. In the Young double-slit experiment, a photon can
``interfere with itself" as a route through one of the slits
superposes with a route through the other slit (actually there are
infinitely many superpositions), resulting in the familiar
interference pattern. It is clear that this does no longer hold
true if the photon is replaced by a nonabelian gauge boson, as the
evolution equations now are nonlinear, destroying the general
superposition possibility. The reason is that the photons can be
represented by harmonic oscillators \cite{Dirac2}, while
nonabelian fields cannot (also casting doubt on the existence of
nonabelian quanta). For example, in a hypothetical world built by
``color" alone, which is not completely absurd, as
``glueball"-states consisting solely of color fields are expected
to exist, the state vector would be constructed from the color
potentials, but as these are described by eight coupled, nonlinear
PDEs, the superposition principle would be lost.

The mathematical formalism underlying all this
can be summarized as follows: \\
For abelian theories, such as QED, a solution to equation (1) can
be found by the usual superposition of Fourier analysis (dropping
the normalization factor),
\begin{eqnarray}
A_{\mu}^{Abelian} = \int d^3 k \sum_{\lambda = 0}^3 [a_k (\lambda)
\epsilon_{\mu}(k, \lambda) e^{-i k \cdot x} + a_k^{\dagger}
(\lambda) \epsilon_{\mu}^{\ast} (k, \lambda) e^{i k \cdot x}],
\end{eqnarray}
where $\epsilon_{\mu}$ is the polarization vector. For a
nonabelian theory this is no longer true,
\begin{equation}
A_{\mu}^{b \, Nonabelian} \neq   \int d^3 k \sum_{\lambda = 0}^3
[a_k^b (\lambda) \epsilon_{\mu}(k, \lambda) e^{-i k \cdot x} +
a_k^{b \, \dagger} (\lambda) \epsilon_{\mu}^{\ast} (k, \lambda)
e^{i k \cdot x}] ,
\end{equation}
as Fourier methods are inapplicable to nonlinear PDEs (see , {\it
e.g.} \cite{Logan}). This signals the breakdown of quantum
mechanical superposition for nonlinear (nonabelian) gauge fields.
As the Fourier coefficients, ($a_k, a_k^{\dagger}$), gives the
connection to the gauge quanta through the corresponding
annihilation and creation operators, the absence of the above
expansion in the nonabelian case puts nonabelian quanta, and hence
adherence to at least naive quantum mechanics, in doubt.

That a quantum mechanical state must be able to ``self-collapse"
in some way is imperative to obtain a ``classical world" of
macroscopic objects, and especially so in quantum cosmology, the
quantum mechanical treatment of the whole universe, where no
``outside" observer exists. The self-induced collapse puts an end
to the infinite regress of quantum superposition, where first the
measuring apparatus obtains a quantum mechanical nature, then the
observer, and so on, ad infinitum, until the whole universe
consists of infinitely many superimposed quantum states, without
any one of them actually being ``realized". The ``many worlds"
interpretation of Everett \cite{Everett} purports to solve this
problem, without collapse, by assuming that we only see events
which take place in one of these branching universes, but it seems
that the fundamental question of \textit{when}, and how, the
universe actually branches is unanswered by that model, this being
the equivalent of the ``measurement problem" in the orthodox
interpretation. There also seems to exist some empirical results
pointing in the direction of self-collapse. For instance, in
\cite{Mohanty}, the experimental evidence of a non-zero
(saturation) decoherence rate for electrons, even at zero
temperature, seems clear. Thus quantum mechanical particles ought
to be able to decohere \textit{intrinsically}, without any
influence from the environment \cite{Zurek} (\textit{e.g.}, ``heat
bath").

\section{Implementation (rough)}
We now turn to a preliminary, admittedly rough implementation of
our idea of self-induced collapse. Following Penrose
\cite{Penrose}, we choose the following expression for the
self-collapse time

\begin{equation}
\tau = \frac{\hbar}{E_{N.L.}},
\end{equation}
where $E_{N.L.}$ is the energy stored in the nonlinear field
configuration of the nonabelian interaction (which in turn depends
on the strength of the coupling). This choice is also compatible
with the experimentally observed saturation temperature in $e$-$e$
interaction corrections for conductivity \cite{Mohanty}. Observe
that the relation is \textit{not} an uncertainty relation, despite
its identical form, as $\tau$ and $E_{N.L.}$ are not
uncertainties. We see that the relation gives the right classical
``correspondence" limit when $\hbar \rightarrow 0$. The build-up
and collapse of the field configuration could follow a pattern
similar to the collapse of other nonlinear waves \cite{Robinson},
making $E_{N.L.}$, and hence $\tau$, effectively random, in
agreement with radioactive decay, etc.

As we want the energy of the full nonlinear theory, but cannot
today calculate this inherently non-perturbative quantity exactly,
we take it to be a characteristic energy for the interaction. If,
for instance for QCD, we as a rough approximation take the energy
to be $E_{N.L.} \sim \Lambda_{QCD} \sim 0.2$ GeV, we get a
``ballpark" figure of $\tau_{QCD} \sim 10^{-23}$s for the collapse
time for \textit{strong} QCD, \textit{e.g.}, inside a
non-disturbed hadron. Although the exact result for $\tau_{QCD}$
might differ by many orders of magnitude, this may help explain
why (semi-)classical models work so well for strong QCD, as the
stronger the interaction is, the more ``classical" it behaves
according to our mechanism.

In our model, the quantum mechanical (linear, unitary) evolution
is constantly punctuated by ``hits" of self-collapse at an average
frequency of $\langle \tau \rangle^{-1}$. This is similar to the
case in orthodox quantum mechanics where an observation, or the
initial preparation of a state, suddenly ``realizes" one of the
potential outcomes, after which the unitary (linear) evolution of
the state takes over until the next observation. A ``macroscopic"
piece of matter has such a huge energy stored in nonlinear field
configurations that $\tau = \hbar / E_{N.L.} \sim 0$,
approximating a continuously collapsing state, \textit{i.e.}, a
classical state. This forbids quantum mechanical effects to
``invade" the macroscopic world, and resolves the
``Schr\"{o}dinger's cat" paradox \cite{Schrodinger} and related
questions such as ``Wigner's friend" \cite{Wigner}, etc.

Note that \textit{any} significant nonlinear interaction, whether
as part of a ``measurement" carried out by conscious beings
\cite{Schrodinger,Wigner} or not, bring about the collapse of
interfering amplitudes into classical states. (In contrast, Wigner
\cite{Wigner} postulated that grossly nonlinear equations of
motion should replace ordinary quantum mechanical evolution for
conscious beings \textit{only}.) Conscious observation is
therefore only a \textit{special case} of the more general
nonlinearity, as all ``measuring apparatuses", including human
beings, consist of both weakly (all constituents) and strongly
(quarks) nonlinearly interacting fields. Hence, there should be no
need to introduce the \textit{mind} into the interpretation of
quantum mechanics at a fundamental level.

If we would consider just pure QED the nonlinear terms would be
absent, hence a hypothetical world built by QED alone would never
be classical. This also explains why, \textit{e.g.}, atomic
physics, and the numerous recent tests of quantum mechanics using
laser setups, works so closely to orthodox quantum mechanics
(before ``measurement"), as it is being ``classicalized" only by
(very) weak interaction effects. Were it not for the existence of
other interactions besides QED, we would indeed have quantum
mechanical superpositions of whole universes, \textit{i.e.}, the
``many worlds" interpretation of quantum mechanics by Everett
\cite{Everett}.

One difference between our proposal for self-induced collapse, and
other models aiming at the same goal, is that, as far as we know,
all other models postulate additional equations and/or variables,
\begin{itemize}
  \item \textit{Decohering histories} \cite{Griffiths,GellMann}: new fundamental
 principle of irreversible coarse graining + additional
constraints to remove ``too many" decoherent histories, also,
superpositions are not really removed \cite{Bell} (different
outcomes still ``coexist" as a mixture after decoherence) so
``measurements" could be undone
  \item \textit{Altered Schr\"{o}dinger equation} \cite{Pearle,GRW}: obvious extra
  (nonlinear) term added
  \item \textit{Bohm QM} \cite{Bohm}: additional (nonlinear) evolution equation
for objective particle positions
\end{itemize}
whereas we use only nonlinearities which are \textit{already
present} in the dynamics of the accepted standard model of
particle physics. Another difference is that, to our knowledge,
all other models for spontaneous collapse/decoherence are
non-relativistic, whereas our scheme is based on covariant
theories. This also opens up for a treatment of ``measurement" in
quantum field theory.

As a nonlinear gauge evolution is effectively non-reversible,
especially if chaotic \cite{Biro}, it lies close to identify it
with the physical ``mechanism" of the ``irreversible
amplification" emphasized by Bohr as being necessary to produce
classical, observable results from the quantum mechanical
formalism. Even though Bohr himself denounced the need, or even
the possibility, to give a physical description of this
``mechanism" \cite{Bohr}, we believe that the central issue for
truly understanding quantum mechanics lies in the quantum
measurement problem. For instance, it is \textit{only} there, in
the collapse of the wave function (or its equivalent), that the
indeterminacy of quantum mechanics enters. It is simply impossible
to obtain the observed random behavior from the linear
Schr\"{o}dinger equation alone, without additional postulates or
assumptions. It may, however, be possible that deterministic chaos
in the nonlinear self-interaction can be responsible for the
seemingly stochastic character of quantum mechanics, by collapsing
to different alternatives effectively at random. In a way, it
would even be surprising if quantum mechanics \textit{is}
fundamentally probabilistic, as in all other cases probabilities
are \textit{derived} from an underlying, deterministic dynamics.
It may even be argued that the statistical patterns that arises in
quantum mechanics may be taken as implicit proof that an
underlying structure is present, as pure randomness would give
\textit{no} statistical correlations, contradicted by,
\textit{e.g.}, the observed definite average lifetimes for
radioactive nuclei or the distinctive pattern arising in a
two-slit experiment. At least one \textit{local}, deterministic
model of this sort has already been constructed \cite{Palmer} in a
somewhat different context (utilizing gravity, and using
``riddled", or intertwined, chaotic attractor basins to sidestep
Bell's theorem \cite{Bell2}).

Our model can be experimentally tested, at least in principle, as
differently charged (electric, weak, color,...) fields should have
different coherence lengths. In a double-slit experiment, for
instance, the photon should have a much longer (in principle
infinite if gravity is ignored) coherence length than,
\textit{e.g.}, electrons which ought to have a finite coherence
length due to nonlinear weak interactions. As the full nonlinear
calculations are very complex, it is not possible to
quantitatively predict the coherence lengths at the present time,
but if it turns out that electrons experimentally have shorter
coherence lengths than photons it would strengthen our
hypothesis\footnote{One referee asked about observed interference
effects in double-slit experiments using neutrons and $C_{60}$.
The results should, according to the ideas presented here, be
$\tau_{\gamma} > \tau_{e} > \tau_{n} > \tau_{C_{60}}$ due to
$\infty = \tau_{pure QED} > \tau_{weak int}
> \tau_{residual QCD}
> \tau_{multinucleon} $. This does not contradict present observational data.}.
As another example of a possible experimental test, the coherent
quantum state of, \textit{e.g.}, the huge number of electrons in a
superconducting ring should slowly decohere by the weak
interaction, destroying the quantum mechanically superposed
magnetic field configurations.

Given that our model is based on somewhat tentative and unfamiliar
concepts, the fact that it should be amenable to experimental
tests of this sort might still give it some credibility. Many of
the other models proposed for solving the measurement problem
predict experimental results identical to orthodox quantum
mechanics, hence making them more interpretive than physical.

The collapse usually postulated in quantum mechanics is not
relativistically covariant, as it is instantaneous over all space.
Only the \textit{deterministic} unitary development of the state
is taken into account by the relativistic Dirac equation and, more
generally, by quantum field theory. As our scheme for collapse is
based on covariant gauge-field theories, it might be possible to
describe the collapse of the state in a covariant way, although
our present rough attempt, which for \textit{simplicity} singles
out the energy accumulated in the nonlinear field configurations,
is not covariant. On the other hand, and perhaps more plausible,
it can be argued that the collapse should be described by an
inherently non-local mechanism, as it seems that quantum mechanics
(and nature) at its very foundation \textit{is} non-local, as
given by the results of Aspect \textit{et al.} \cite{Aspect}, and
more recent experiments on quantum non-separability. As a
nonlinear theory also in some cases can be non-local, it would be
interesting to investigate if this view of spontaneous collapse
can account for these crucial non-local effects. Work in this
direction is in progress \cite{Dugne}, together with a more
detailed investigation of the nonlinear terms in the nonabelian
evolution equations, as a means to better understand the
quantitative details of the proposed mechanism for self-collapse.

\section{Conclusion}
The gist of our proposal can be illustrated by an analogy:

i) Do ocean waves collapse (``break") only when they are observed?
No, they automatically collapse as dictated by their nonlinear
dynamics.

ii) Do quantum mechanical waves collapse only when observed? No,
they automatically collapse as dictated by the nonlinear dynamics
in their interactions (gauge fields).

The analogy admittedly halts a little, as quantum mechanics
``lives" in an abstract, multidimensional configuration space.

However, it is a fact that
\begin{itemize}
\item no consistent solution to the quantization of strongly
nonlinear theories yet exists
\item nonlinear terms break quantum
mechanical superposition
\end{itemize}
Maybe nature is trying to tell us that strongly nonlinear theories
should \textit{not} be quantized, and that they instead are
responsible for turning the quantum ``potentialities" into
classical ``actualities"?

We have described how automatic dynamical collapse of the wave
function in quantum mechanics may already be implicit in the
existing dynamical theory, that of nonabelian gauge fields. These
include the weak interaction, QCD, gravity, and any other
nonabelian fields which eventually may be found in the future. The
nonlinear self-interaction terms break the fundamental
superposition principle of quantum mechanics, inducing just the
right physical mechanism for the purpose of solving the
``measurement problem".

\end{document}